\documentclass[aps,prl,twocolumn,showpacs,floatfix]{revtex4}

\usepackage{graphicx,amsmath,bm}

\usepackage{amssymb}
\usepackage{amsfonts}
\usepackage{mathtools}

\usepackage{placeins}

\newcommand{\ii}{\mathrm{i}}
\newcommand{\Ca}{CaFe$_2$As$_2$}
\newcommand{\PCa}{Ca(FeAs$_{0.875}$P$_{0.125}$)$_2$}

\newcommand{\vect}[1]{{\bm #1}}
\newcommand{\mat}[1]{\hat{#1}}
\newcommand{\vectmat}[1]{{\hat{\bm{#1}}}}
\newcommand{\tr}{\mathrm{tr}}

\DeclarePairedDelimiter{\bra}{\langle}{\rvert}
\DeclarePairedDelimiter{\ket}{\lvert}{\rangle}

\newcommand{\braket}[2]{\left\langle #1 \vphantom{#2} \right.
  \left\lvert #2 \vphantom{#1} \right\rangle}

\newcommand{\ham}{\mat{H}}
\newcommand{\lattice}{\mathcal{L}}

\newcommand{\bzlatt}{\mathcal{B}(\lattice)}

\newcommand{\vecn}{\vect{n}}
\newcommand{\vecm}{\vect{m}}

\newcommand{\siten}{\vect{R}_{\vecn}}
\newcommand{\sitem}{\vect{R}_{\vecm}}

\newcommand{\transn}{\mat{T}_{\vecn}}
\newcommand{\transm}{\mat{T}_{\vecm}}

\newcommand{\transNi}{\mat{T}_{N_i}}
\newcommand{\transai}{\mat{T}_{\vect{a}_i}}
\newcommand{\gtrans}{\mathbb{T}}

\newcommand{\irrep}[1]{\Delta^{(#1)}}
\newcommand{\irrepel}[2]{\Delta^{(#1)}(#2)}
\newcommand{\irrepelc}[2]{\big[\Delta^{(#1)}(#2)\big]^\dagger}

\newcommand{\kvec}{\vect{k}}

\newcommand{\dvec}{\vect{k}_f^{\alpha\beta}}

\newcommand{\projk}{\mat{P}_{\kvec}}
\newcommand{\subspk}{\mathbb{V}_{\kvec}}
\newcommand{\hamk}{\ham_{\kvec}}
\newcommand{\hamka}{\ham_{\kvec\alpha}}
\newcommand{\hamkb}{\ham_{\kvec\beta}}

\newcommand{\excop}{\vect{\varepsilon\cdot\vectmat{x}}}

\newcommand{\muparam}[1]{#1,\mu}
\newcommand{\nuparam}[1]{#1,\nu}

\newcommand{\ketmu}[1]{\ket{\muparam{#1}}}
\newcommand{\ketnu}[1]{\ket{\nuparam{#1}}}
\newcommand{\bramu}[1]{\bra{\muparam{#1}}}

\newcommand{\musite}{\vect{s}_{\mu}}

\newcommand{\ketk}[1]{\ket{\kvec,#1}}
\newcommand{\brak}[1]{\bra{\kvec,#1}}

\newcommand{\sgroup}{\mathbb{S}}

\newcommand{\cosop}{\mat{C}}
\newcommand{\cosopi}{\cosop_i}

\newcommand{\sgopi}{\mat{S}_{\vecn i}}

\newcommand{\pointop}{\mat{U}}
\newcommand{\fractrans}{\vect{\tau}}
\newcommand{\pointopi}{\pointop_i}
\newcommand{\fractransi}{\fractrans_i}

\newcommand{\seitzop}[2]{(#1|#2)}

\newcommand{\seitzcosi}{\seitzop{\pointopi}{\fractransi}}

\newcommand{\seitzsgni}{\seitzop{\pointopi}{\siten+\fractransi}}

\newcommand{\rvec}{\vect{r}}

\newcommand{\littleg}{\mathbb{L}_{\kvec}}
\newcommand{\pointki}{\mat{U}_i^{\kvec}}

\newcommand{\Kvec}{\vect{K}}

\newcommand{\lirrepa}{\Lambda^{(\kvec,\alpha)}}

\newcommand{\gammasite}{\vect{s}_{\gamma}}
\newcommand{\deltasite}{\vect{s}_{\delta}}

\newcommand{\projka}[1]{\mat{P}_{\kvec #1}}

\newcommand{\factgroup}{\mathbb{C}}

\newcommand{\fia}{f_i^\alpha}

\newcommand{\fib}{f_i^\beta}

\newcommand{\Ir}{Irreducible representation}
\newcommand{\ir}{irreducible representation}
\newcommand{\irs}{irreducible representations}

\begin{document}

\title{Unfolding of the electronic structure through the induced
representations of space groups: Application to Fe-based superconductors}

\author{Milan Tomi{\'c}}
\email[Email:]{tomic@itp.uni-frankfurt.de}
\affiliation{Institut f\"ur Theoretische Physik, Goethe-Universit\"at Frankfurt,
Max-von-Laue-Stra{\ss}e 1, 60438 Frankfurt am Main, Germany}

\author{Harald O. Jeschke}

\affiliation{Institut f\"ur Theoretische Physik, Goethe-Universit\"at Frankfurt,
Max-von-Laue-Stra{\ss}e 1, 60438 Frankfurt am Main, Germany}

\author{Roser Valent\'\i}

\affiliation{Institut f\"ur Theoretische Physik, Goethe-Universit\"at Frankfurt,
Max-von-Laue-Stra{\ss}e 1, 60438 Frankfurt am Main, Germany}

\date{\today}

\begin{abstract}
We revisit the problem that relevant parts of bandstructures for a given  cell
choice can reflect exact or approximate higher symmetries of subsystems in the
cell and can therefore be significantly simplified by an unfolding procedure
that recovers the higher symmetry.  We show that bandstructure unfolding can be
understood as projection onto induced {\irs} of a group obtained by extending
the original group of translations with a number of additional symmetry
operations. The resulting framework allows us to define a generalized unfolding
procedure which includes the point group operations and can be applied to any
quantity in the reciprocal space. The unfolding of the Brillouin zone follows
naturally from the properties of the induced {\irs}. In this context, we also
introduce a procedure to derive tight-binding models of reduced dimensionality
by making use of point group symmetries. Further, we show that careful
consideration of unfolding has  important consequences on the interpretation of
angle resolved photoemission experiments. Finally, we apply the unfolding
procedure to various representative examples of Fe-based superconductor
compounds and show that the one iron picture arises as an irreducible
representation of the glide mirror group and we comment on the consequences
for the interpretation of one-iron versus two-iron Brillouin zone representations.
\end{abstract}


\maketitle

\section{Introduction}

{\it Ab initio} electronic structure calculations have become a primary tool of
research for understanding the microscopic behavior of solids.  The multitude of
methods that have emerged to deal with periodic crystal systems, such as density
functional theory (DFT)~\cite{Hohenberg1964}, rely on the Bloch
theorem\cite{Bloch1928} in one form or the other. The Bloch theorem builds upon
the translational symmetry of the crystal lattice and paves the way to the
fundamental concepts needed to understand the electronic structure of periodic
systems, such as the classification of the electronic states in terms of
wavevectors $\kvec$ and the notion of bandstructure. Many important properties
of the crystal, such as magnetic or transport properties, are encoded in the
bandstructure. In addition, the bandstructure is important for the
interpretation of a few experimental measurements like angle resolved
photoemission spectroscopy (ARPES).

However, a problem arises in the practical use of DFT calculations  whenever we
have to deal with  systems where the original translational symmetry is broken.
These situations are encountered, for example, in calculations on  doped
materials or in magnetically ordered systems. Often, in these situations we
have to employ large supercells whose size determines the periodicity of the
bandstructure through the Bloch theorem. This results in a complicated
bandstructure consisting of many bands, which is hard to interpret.

Recently, a number of methods have emerged \cite{Boykin2005, Boykin2007,
Boykin2007_2, Ku2010, Popescu2012, Medeiros2014, Huang2014, Chen2014} to
alleviate this problem. A common approach shared among these methods, implicitly
or explicitly, is  a transformation from one Bloch basis to another.

An important aspect of the Bloch theorem is that it is an expression of one of
the fundamental group-theoretical principles, which states that the eigenstates
of a physical system can be classified according to the irreducible
representations of its group of symmetries~\cite{Wigner1931}. In light of this,
the bandstructure unfolding can be viewed as a transformation between the {\irs}
of different translation groups. Despite the number of recent publications
~\cite{Boykin2005, Boykin2007, Boykin2007_2, Ku2010,Popescu2012, Medeiros2014,
Huang2014}, a rigorous consideration of the group-theoretical aspects of the
bandstructure unfolding is still missing and as a consequence certain important
properties are overseen, especially in relation to ARPES experiments.

The purpose of this work is to attempt to close this conceptual gap by
introducing a bandstructure unfolding based on group theory. This treatment 
allows us to incorporate point group symmetries into a unified framework and
generalizes the idea of using glide-mirror operations, initially proposed in the
context of LaFeAsO~\cite{Andersen2011}, in order to obtain models of reduced
dimensionality. Our bandstructure unfolding also allows for a clear
understanding of the two-Fe versus one-Fe description of the electronic
properties of Fe-based superconductors.

We will show that bandstructure unfolding can be achieved by projecting the
bands onto the induced {\irs} of the supergroup of the initial group of
translations. We will also show that this leads naturally to the concept of the
unfolded Brillouin zone. With the help of the point group operations,
bandstructures can be unfolded beyond the limits of  translational symmetry.
Further, tight-binding models with  reduced number of orbitals can be formulated
under certain conditions.

The group-theoretical formulation of the unfolding procedure in terms of
projections onto the irreducible subspaces allows us to unfold any quantity in
the reciprocal space if we know how it behaves under the symmetry operations of
the crystal lattice. In addition, the unfolding artifacts such as
``ghost-bands'' \cite{Ku2010} or ``incomplete bands'' \cite{Lin2011} are
naturally explained as bands with projections onto the multiple irreducible
subspaces in cases with broken symmetry.

\section{Method}

\subsection{Group of translations}

One of the fundamental statements in solid state physics is that the eigenstates
of the Hamiltonian of a periodic system can be classified according to the
{\irs} of the group of translational symmetries of the system.

Let us denote the crystal lattice as \mbox{$\lattice = \{\siten =  \sum_{i=1}^3
n_i\vect{a}_i |n_i=1,...,N_i\}$}, where $\vecn=(n_1, n_2, n_3)$, $N = N_1 N_2
N_3 $ is the number of unit cells in the lattice, $\siten$ are their position
vectors, and $\vect{a}_i$ are the vectors spanning the unit cell. The lattice is
invariant under the action of the group of translation operators $\gtrans =
\{\transn\}$, such that $\siten+\sitem=\transm\siten$, where periodic boundary
conditions $\transNi=1$ are assumed. The translation group $\gtrans$ is an
Abelian cyclic group generated by the three generators $\transai$. As such, its
{\irs} are one-dimensional and given by $\irrepel{\kvec}{\transn} =
\exp(-\ii\kvec\cdot\siten)$. There are $N$ inequivalent {\irs} of $\gtrans$ and
they can be enumerated by the vector index $\kvec\in\bzlatt$, where $\bzlatt =
\{\sum_{i=1}^3 g_i\vect{b}_i | g_i\in[0, 1)\}$ is the Brillouin zone (BZ), and
$\vect{b}_i=2\pi \varepsilon_{ijk}\vect{a}_j\times\vect{a}_k /\big[\vect{a}_1
\cdot (\vect{a}_2\times\vect{a}_3)\big]$ are unit vectors of the reciprocal
lattice $\lattice^{-1}$. To each {\ir} $\irrep{\kvec}$  corresponds a one
dimensional subspace $\subspk$,
defined as a co-domain of the projection operator
\begin{equation}\begin{split}\label{eq:trans_projector}
\projk &= \frac{1}{\sqrt{N}}\sum_{\vecn}\irrepelc{\kvec}{\transn} \transn \\&=
\frac{1}{\sqrt{N}}\sum_{\vecn}\exp(\ii\kvec\cdot\siten)\transn
\end{split}\end{equation}
Because for all $\vecn$, $[\transn,\ham]=0$,  the subspaces $\subspk$ will be
orthogonal irreducible subspaces of the Hamiltonian $\ham$, and thus, a symmetry
classification of the eigenstates of the Hamiltonian is achieved.

In order to proceed, we assume that we have $P$ localized electronic states,
$\ketmu{\vect{0}}$, $\mu=1,\dots,P$, centered at positions $\musite$, occupying
the unit cell located at origin $\vect{0}$. The sites $\musite$ don't
necessarily have to be different since we can consider cases with multiple
orbitals per atomic site. The translationally invariant electronic states of the
crystal lattice are obtained by the action of $\gtrans$ onto these states, and
the resulting localized electronic states are $\ketmu{\siten} = \transn
\ketmu{\vect{0}}$ at positions $\siten+\musite$. Application of the projector
\eqref{eq:trans_projector} to the states $\ketmu{\vect{0}}$ results in the
familiar Bloch-states 
\begin{equation}\label{eq:bloch_basis}
 \ketk{\mu} = \frac{1}{\sqrt{N}}\sum_{\vecn} \exp(\ii\kvec\cdot\siten)
\ketmu{\siten}
\end{equation}
Using \eqref{eq:bloch_basis} as a basis, the Hamiltonian $\ham$ is brought to
the block-diagonal form, with $N$ blocks of size $P\times P$, whose elements are
given by $[\hamk]_{\mu\nu} = \brak{\mu}\ham\ketk{\nu}$. Every block $\hamk$ can
be diagonalized separately, yielding a set of $P$ bands at $\kvec$. A set of
bands for every $\kvec\in\bzlatt$ represents the bandstructure for the given
lattice $\lattice$.

\subsection{Extension to additional symmetry operations}

In many applications, the translation group $\gtrans$ can be expanded with a
certain number of operations, which are, approximately, symmetry operations of
the lattice $\lattice$. When this is the case, we will have a new group
$\sgroup$, such that $\gtrans$ is an invariant subgroup of $\sgroup$, denoted
$\gtrans \triangleleft \sgroup$. Owing to this fact, {\irs} of $\sgroup$ can be
induced from $\gtrans$ in a simple manner. Furthermore, addition of every new
symmetry operation will halve the number of independent states $\ketk{\mu}$ at
every $\kvec$, since additional operations will produce one half of the states
$\ketk{\mu}$ from another half, and thus will halve the number of bands,
producing the \emph{unfolded} bandstructure. This is subject to certain
conditions which will be outlined in the discussion that follows.

Let us assume now that we are expanding $\gtrans$ with $K$
operations $\cosopi$,
denoted in Seitz notation~\cite{ElBatanouny2008,Kim1999} as $\cosopi = \seitzcosi$, $i=1,..,K$, 
 where
$\pointopi$ is a point group operation and $\fractransi$ is a fractional
translation (with respect to the translations $\transn$ of $\gtrans$) so that
the combination of $\cosopi$ and $\transn$ leads to a space group $\sgroup$.
Operations $\cosopi$ are allowed to be pure translations but not pure point
group operations, since in the case of the pure point group operations, we
would not be able to interpret the unfolded bandstructure in terms of a lattice
of reduced periodicity. 

The action of $\cosopi$ on an arbitrary point in the Cartesian space is given by
\begin{equation*}
 \cosopi\rvec=\pointopi\rvec+\fractransi
\end{equation*}
while the combined action of $\cosopi$ and $\transn$ results in the space group
operation $\sgopi$ defined as
\begin{equation*}
 \sgopi\rvec=\transn\cosopi\rvec=\pointopi\rvec+\siten+\fractransi = 
  \seitzsgni\rvec
\end{equation*}
The space group operators $\sgopi$ induce an action on the localized states.
Under the point group operations $\pointopi$, and fractional translations
$\fractransi$, states $\ketmu{\siten}$ transform into each other as
$\seitzcosi\ketmu{\siten}=\sum_{\nu}\ketnu{\siten} W_{\nu\mu}(\cosopi)$, where
the matrices $\mat{W}(\cosopi)$ represent the action of the operations
$\cosopi=\seitzcosi$ in the basis composed of the states $\ketmu{\siten}$. Their
matrix elements can be written as $W_{\gamma\delta}(\cosopi) = r_{\gamma\delta}
\delta(\deltasite - \cosopi\gammasite)$, where $\delta$ is the Kronecker delta
and sites $\deltasite$ and $\cosopi \gammasite$ are considered equal if they
differ by a lattice vector. The total action of $\sgopi$ is
\begin{equation}
 \sgopi\ketmu{\sitem}=\sum_{\nu}\ketnu{\sitem+\siten}
  W_{\nu\mu}(\cosopi)
\end{equation}
The task is now to induce the {\irs} of the space group $\sgroup$. We first
note that the operators $\cosopi$ are the right coset representatives of
$\sgroup$ with respect to $\gtrans$. Let the factor group, corresponding to
this right coset decomposition be $\factgroup=\sgroup:\gtrans$. The well known
property of space groups is that every space group is solvable, that is, every
space group can be decomposed into a series $\sgroup_0 \triangleleft \sgroup_1
\triangleleft ... \triangleleft\sgroup_D = \sgroup$, where every factor group in
the decomposition is Abelian. In addition to that, it is always possible to find
the decomposition series where the factor groups $\factgroup_i = \sgroup_{i+1} :
\sgroup_{i}$ are cyclic groups of index two or three. This simplifies the
induction procedure further, since in case the factor group $\factgroup$ is not
a cyclic group, we can always decompose it into a subgroup series $\factgroup_0
\triangleleft \factgroup_1 \triangleleft ... \triangleleft \factgroup_B =
\factgroup$, where every $\factgroup_i$ is a cyclic group.

\begin{figure}
\begin{center}
 \includegraphics{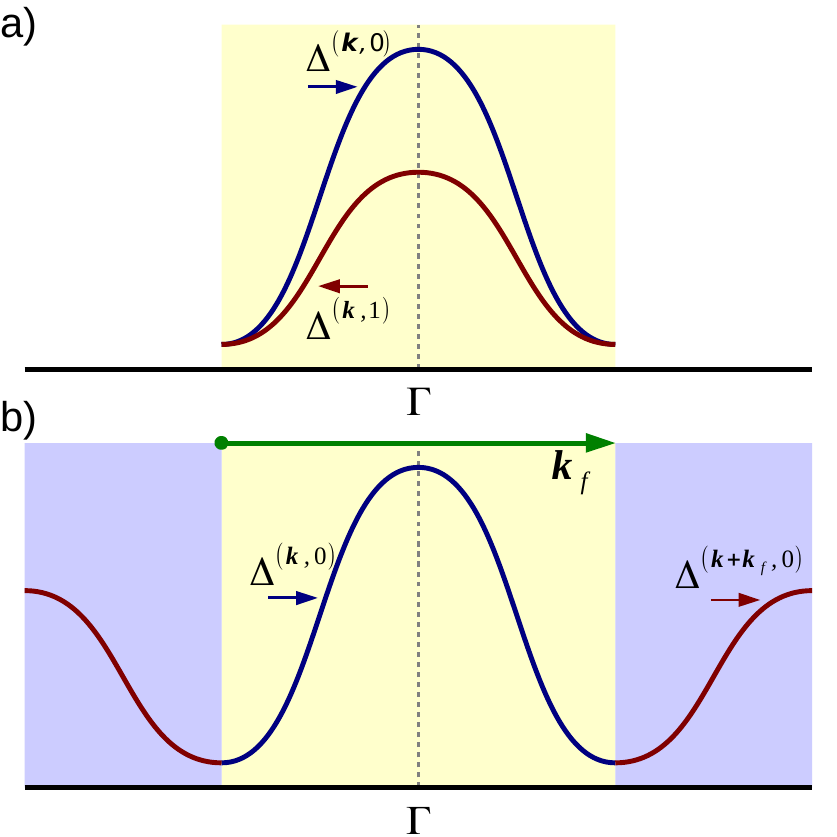}
\end{center}
\caption{Brillouin zone unfolding. (a) Folded bandstructure showing two bands
belonging to two irreducible representations $\irrep{\kvec,0}$ and $\irrep{\kvec,1}$ of $\sgroup$.
(b) The unfolded bandstructure shows that $\irrep{\kvec,1}=\irrep{\kvec +
\kvec_f,0}$, leading to the unfolded Brillouin zone, shown in yellow and purple
backgrounds. The folding vector $\kvec_f$ is shown in green color.}
\label{fig:bz_unfolding}
\end{figure}

In the induction procedure, from every {\ir} of $\gtrans$, multiple {\irs} of
$\sgroup$ can be induced. Some of these {\irs} will be one-dimensional, while
some will be multi-dimensional. The multi-dimensional induced {\irs} of
$\sgroup$ will mix {\irs} of $\gtrans$ with different $\kvec$. Since our goal
is to perform the unfolding within the same $\kvec$, we have to restrict
ourselves only to the cases where the induction procedure yields one-dimensional
{\irs} of $\sgroup$.

\subsection{One-dimensional irreducible representations of $\sgroup$.
Unfolding procedure}

For every {\ir} $\irrep{\kvec}$ of $\gtrans$, {\irs} of $\sgroup$ are
determined from the little co-group $\littleg$~\cite{ElBatanouny2008,Kim1999}. The little co-group is a
group of all point group operations $\pointki$ such that $\pointki\kvec = \kvec
+ \Kvec$, where $\Kvec$ is a reciprocal lattice vector. Given an {\ir}
$\lirrepa$ of $\littleg$, an {\ir} $\irrep{\kvec,\alpha}$ of $\sgroup$ is
induced, where the coset representatives $\cosopi$ are represented by
\begin{equation*}
 \irrepel{\kvec,\alpha}{\seitzcosi}= \exp(-\ii\kvec\cdot\fractransi)
\lirrepa(\pointopi)
\end{equation*}
and the index $\alpha$ runs over all irreducible representations.
This holds across the entire interior of the BZ, with the exception of
the BZ boundary for the cases of the symmorphic space groups, where
the induction procedure is more complex, and more sophisticated
methods, such as Herring's method, are
needed~\cite{ElBatanouny2008,Kim1999}. We will just assume that
$\kvec$ never lies on the BZ boundary, but it can be arbitrarily close
to it. In cases where $\pointopi = 1$, i.e. where the operators
$\cosopi$ are just fractional translations, which is the case of
translational unfolding, the little group $\littleg$ will contain all
fractional translations and will be the same for every $\kvec$. In
this case, the {\irs} of the little group $\lirrepa$, are taken to be
the {\irs} of the group of fractional translations, modulo $\gtrans$,
meaning that two fractional translations are considered to be
identical if they differ by $\siten$.

When $\kvec$ is invariant (up to the reciprocal lattice vector $\Kvec$) under
all point group operators of $\factgroup$, then from the {\ir} $\irrep{\kvec}$
of $\gtrans$, $K$ one-dimensional {\irs} $\irrep{\kvec,\alpha}$ will be induced.
When this is the case, the $\lirrepa(\pointopi)$ will be roots of unity so that
we can write, in general,
\begin{equation}\label{eq:allowed_reps}
 \irrepel{\kvec,\alpha}{\seitzcosi} = \exp(-2\pi\ii\fia/K)
  \exp(-\ii\kvec\cdot\fractransi)
\end{equation}
where $\fia$ is an integer such that $0\leq\fia <K$. The integers $\fia$, with
the operation of addition modulo $K$, constitute a group isomorphic to
$\littleg$. We use the convention that $\alpha=0$ denotes the unit {\ir}, thus
$f_i^0=0$. When \eqref{eq:allowed_reps} is taken into account, the projectors
onto the irreducible subspaces are given by
\begin{equation*}
 \projka{\alpha}=\frac{1}{\sqrt{KN}}\sum_{\vecn}\sum_i
  \exp(\ii\kvec\cdot\siten)\irrepelc{\kvec,\alpha}{\cosopi}\sgopi
\end{equation*}
so that the Bloch basis \eqref{eq:bloch_basis} corresponding to $\sgroup$ can be
written, in analogy to \eqref{eq:trans_projector}, as
\begin{widetext}
\begin{equation}\label{eq:sg_bloch_basis}
 \ketk{\mu,\alpha} = \frac{1}{\sqrt{KN}}\sum_\vecn\sum_i\sum_{\nu}
  \exp(2\pi \ii\fia/K)\exp\big(\ii\kvec\cdot(\siten+\fractransi)\big)
  \ketnu{\siten}W_{\nu\mu}(\cosopi)
\end{equation}
\end{widetext}
By employing the basis Eq.~\eqref{eq:sg_bloch_basis}, $\hamk$ can be brought
into block-diagonal form with $F$ blocks $\hamka$ of size $P/K\times P/K$.  Each
block $\hamka$ can be diagonalized separately and will yield $P/K$ bands. It is
important to note that because of the addition of a fractional translation in
Eq.~\eqref{eq:sg_bloch_basis}, the reciprocal space period of blocks $\hamka$ is
larger than the BZ. Since the only difference between the blocks $\hamka$ is in
the exponential prefactor $\exp(-2\pi\ii\fia/K)$ in \eqref{eq:sg_bloch_basis},
we can restrict ourselves to only one block, i.\,e. $\hamka$, and then reproduce
the other blocks, by allowing  $\kvec$ to leave the BZ, because we can choose
$\dvec\in\bzlatt$, such that 
\begin{align}\label{eq:folding_vecs}
\begin{split}
\dvec\cdot\siten &= 2\pi z \\
\dvec\cdot\fractransi &= 2\pi\frac{\fib-\fia}{K}
\end{split}
\end{align}
where $z$ is an integer. In this way, we can
have $\hamkb=\ham_{\kvec+\dvec,\alpha}$. The vectors $\dvec$ are the
\emph{folding} vectors. The diagonalization of the block $\hamka$ and its
extensions outside of the BZ, by the folding vectors, produces the
\emph{unfolded} bands. The BZ, enlarged by the $K$ folding vectors $\dvec$ is
the unfolded BZ (Fig.~\ref{fig:bz_unfolding}).

It is important to stress here, that due to the fact that $\cosopi$ involve the
fractional translations $\fractransi$, and that due to the requirement of the
one-dimensionality of the {\irs} of $\sgroup$, the unfolded bandstructure
represents the bandstructure of the crystal lattice with the unit cell size
reduced by a factor of $K$, which can be simply translationally folded back
along the folding vectors to represent the starting bandstructure, regardless of
the point group operations $\pointopi$. Because of this, we can effectively
describe the electronic structure with a Hamiltonian of smaller
dimensionality. We can also understand this in a different way. Since
$\ketk{\mu,\alpha}$ are the symmetry adapted basis vectors, no interaction
contained in the Hamiltonian can cause a transition between the states with
different $\alpha$, thereafter all dynamical processes are contained within
their respective irreducible subspaces. We have seen that different {\irs}
become equivalent if shifted by the folding vectors, meaning that no information
is lost if we just keep a single {\ir}, as long as we expand it onto the entire
unfolded Brillouin zone.

One more important issue to note is that our requirement for the one-dimensional
{\irs} of $\sgroup$, implies that the unfolding which utilizes the point group
operations is exact only for $\kvec$ values which are invariant under all point group
operations $\pointopi$. For example, if we use the screw-axis operations, the
unfolding will be exact only along a corresponding high-symmetry line in the
Brillouin zone. However, if the electronic properties are dominantly one
dimensional along the given high-symmetry line, the unfolding can still be used
across the entire Brillouin zone, while preserving the accuracy up to a
significant degree.

To unfold the bandstructure given in the Bloch basis \eqref{eq:bloch_basis}, we
just need to calculate the matrix elements of the projectors $\projka{\alpha}$.
The matrix elements are given by
\begin{multline}\label{eq:projectors}
[\projka{\alpha}]_{\gamma\delta} = 
  \brak{\gamma} \projka{\alpha} \ketk{\delta} \\=
  \frac{1}{K}\sum_i
  \exp(2\pi \ii\alpha\fia/K)
  \exp(\ii\kvec\cdot\fractransi)W_{\gamma\delta}(\cosopi)
\end{multline}
If we assume that from the bandstructure calculations we obtain bands
$\ketk{n}$, where $n$ is the band index, we can unfold the bands by
applying the projectors \eqref{eq:projectors} to the column-vector
containing the projections of bands onto the localized states
$w_{\kvec,n}^{\mu} = \braket{\vect{0},\mu} {\vect{k},n}$. In our
particular case, we have used the Vienna Ab-Initio Simulations Package
(VASP) to obtain the bandstructure. Since, within the VASP package,
the exponential factors $\exp(-i\kvec\cdot\musite)$ for fractional
site vector $\musite$ are already included in the projections
$w_{\kvec,n}^{\mu}$, the exponential factors in
Eq.~\eqref{eq:projectors} can be omitted, simplifying the expressions
even further.

\subsection{Unfolding of tight-binding models}
In general, any observable $\mat{A}_{\kvec}$ can be unfolded by employing the
projectors $\projka{\alpha}$ so that $\mat{A}_{\kvec\alpha} = \projka{\alpha}
\mat{A}_{\kvec} \projka{\alpha}$. 

With the help of Eq.~\eqref{eq:sg_bloch_basis}
 it is also possible to unfold tight-binding  models.
 The matrix elements of the Hamiltonian in the tight-binding
model are defined in the Bloch basis Eq.~\eqref{eq:bloch_basis} as
\begin{equation}\begin{split}\label{eq:tb_model}
 [\hamk]_{\mu\nu} &= 
  \bramu{\vect{0}} \mat{H}\projk \ketnu{\vect{0}} \\ &=
  \sum_{\siten} t_{\mu\nu}(\siten)\exp(\ii\kvec\cdot\siten)
\end{split}\end{equation}
where $t_{\mu\nu}(\siten) = \bramu{\vect{0}}\mat{H}\ketnu{\siten}$ are
the hopping energies. This result follows from $\hamk = \projk\ham\projk$ and
the facts that $\ham$ commutes with $\projk$ and that $\projk$ is idempotent.
We define the  matrix composed of the hopping energies $t_{\mu\nu}(\siten)$ as
$\mat{t}(\siten)$. The unfolding of the tight-binding
 model is achieved by calculating the
matrix elements of the Hamiltonian in the basis Eq.~\eqref{eq:sg_bloch_basis} 
and
then casting the resulting expression in the form of Eq.~\eqref{eq:tb_model}
\begin{equation}\begin{split}\label{eq:sg_tb_model}
[\hamka]_{\mu\nu} &= 
  \bramu{\vect{0}} \mat{H}\projka{\alpha} \ketnu{\vect{0}} \\&=
  \sum_{\vecn}\sum_i
  t_{\mu\nu}^\alpha(\siten+\fractransi)
  \exp(\ii\kvec\cdot(\siten+\fractransi))
\end{split}\end{equation}
where the corresponding hopping energy matrix is $\mat{t}^{\alpha}(\siten +
\fractransi)$. The hopping energies of the unfolded TB model can then be read
off as coefficients of the exponential terms $\exp(\ii\kvec \cdot (\siten +
\fractransi))$. The hopping energy matrices $\mat{t}^{\alpha}(\siten +
\fractransi)$ will have the same block-diagonal structure of $\hamka$. The
general expression for the hopping energies is then
%
%\begin{widetext}
\begin{equation}\label{eq:unfolded_hoppings}
 \mat{t}^{\alpha}(\siten+\fractransi) = 
  \frac{\exp(2\pi\ii\fia/K)}{K}\mat{t}(\siten)\mat{W}(\cosopi)
\end{equation}
%\end{widetext}
%
It should be noted here that for $K>2$, the unfolded hopping energies can
become complex due to the prefactor $\exp(2\pi\ii\fia/K)$. However, this
prefactor does not affect the eigenvalues and eigenvectors of $\hamka$ since it
amounts to an overall, $\kvec$-independent, unitary transformation of the
Hamiltonian. Furthermore, in the unfolded picture, we extend
 a single {\ir} beyond
the BZ boundaries, so that in the tight-binding model
 we expect to have a single
set of hopping energies independent of the {\ir}.  This
 manifestly does not
hold for Eq.~\eqref{eq:unfolded_hoppings} since the exponential prefactor depends
on $\alpha$. Because of this, the exponential prefactor in
Eq.~\eqref{eq:unfolded_hoppings} can be dropped and the {\ir}-independent unfolded
hopping energies can be defined as
\begin{equation}\label{eq:unfolded_hoppings_final}
 \mat{t}(\siten+\fractransi) = \frac{1}{K}\mat{t}(\siten)\mat{W}(\cosopi)
\end{equation}
When the unfolded hopping energies are defined in this way, the index
$\alpha$ can be omitted from Eq.~\eqref{eq:sg_tb_model}. Since we
already concluded that the $\mat{t}^{\alpha}(\siten+\fractransi)$ are
block-diagonal, a reduction of the dimensionality of the tight-binding
model is achieved. Practically, this means that orbital indices $\mu$
and $\nu$ can be taken to run only over the first block in the
block-diagonalized Hamiltonian, since other blocks are symmetrically
equivalent.

\subsection{Relation to ARPES}

Angle resolved photoemission is one of the most direct ways to experimentally
observe the bandstructure in solids~\cite{Damascelli2004}. However, the
interpretation of raw experimental data is a very complicated process and often
relies heavily on comparisons with density functional theory calculations. This
becomes especially difficult in systems with broken translational symmetry,
since, on the one hand, supercell calculations have to be employed by density
functional theory resulting in complicated folded bandstructures, while on the
other hand ARPES data often shows the unfolded bandstructure, sometimes offset
away from the first Brillouin zone~\cite{Ku2010,Lee2010}. This was already
discussed in the context of bandstructure unfolding~\cite{Ku2010}. However, some
important issues were not considered, like the fact that multiple {\irs} are
involved in the unfolding as well as the effect of high symmetry elements of the
Brillouin zone.

The observed photoelectron intensity at energy $\omega$ in ARPES experiments can
be directly related to the one-electron spectral function~\cite{Borstel1985,
	Nolting1990}
\begin{equation*}
I(\omega) \sim \sum_{\kvec}\sum_f\sum_{ij}
M_{\kvec fi} A_{\kvec ij}(\omega) M_{\kvec jf}
\end{equation*}
where $M_{\kvec fi} = \brak{f}\excop\ketk{i}$ are the dipole matrix elements
between the initial state $\ketk{i}$ and the final state $\ketk{f}$ and
$\mat{A}_{\kvec ij}(\omega)$ are the matrix elements of the one-electron
spectral function. Using the projectors from Eq.~\eqref{eq:trans_projector} this can
be rewritten as
\begin{equation}\label{eq:arpes_intensity}
I(\omega) \sim 
\tr\left(\sum_{\kvec}\excop\projk\mat{A}(\omega)\projk\excop\right)
\end{equation}
where the trace is taken only over the final states and $\mat{A}$ is restricted
to the occupied subspace. If the observed system has approximate symmetry
given by a group $\sgroup\triangleright\gtrans$, either because
(i) we have a slight breaking of the translational symmetry, or 
(ii)  most of the
contribution to the sum over $\kvec$ in Eq.~\eqref{eq:arpes_intensity}
comes from
the surface states lying in the high-symmetry element of the Brillouin zone, we
can extend the summation over $\kvec$ onto the irreducible subspaces of
$\sgroup$ so that we have
\begin{equation*}
I(\omega) \sim \tr\left(\sum_{\kvec} \sum_{\alpha \beta}
\excop\projka{\alpha}\mat{A}(\omega)\projka{\beta}\excop\right)
\end{equation*}
Since $\sgroup$ is an approximate symmetry, the off-diagonal blocks
$\projka{\alpha} \mat{A} \projka{\beta}$ can be neglected. In
addition, we can use the folding relations,
Eq.~\eqref{eq:folding_vecs} to replace the summation over the {\ir}
index $\alpha$ with the summation over the unfolded Brillouin zone to
obtain
\begin{equation}\label{eq:arpes_intensity_final}
I(\omega) \sim \tr\left(\sum_{\kvec_{\alpha}}
\excop\projka{\alpha}\mat{A}(\omega)\projka{\alpha}\excop\right)
\end{equation}
where $\kvec_{\alpha}=\kvec+\dvec, \forall\beta$, belong to the
Brillouin zone unfolded in accordance with
Eq.~\eqref{eq:folding_vecs}. In this case, the ARPES experiment will
observe the spectral function $\mat{A}_{\kvec\alpha}(\omega) =
\projka{\alpha} \mat{A}(\omega) \projka{\alpha}$ instead of the
spectral function $\mat{A}_{\kvec}(\omega) = \projk \mat{A}(\omega)
\projk$. One important detail in Eq.~\eqref{eq:arpes_intensity_final}
is the fact that the {\ir} $\alpha$ is only selected through the
effect of dipole operators and the trace over the final states. This
means, that the actual {\ir} observed in the experiment depends on the
experimental conditions.

This is a very important conclusion, since it tells us that in order
to carefully interpret raw ARPES data, we have to take into account
two considerations.  The first one is the possible effect of high
symmetry elements in the Brillouin zone. This implies the
incorporation of point group operations into the unfolding -- we would
like to stress here, that this issue is distinct from orbital symmetry
selection by means of polarization. The second consideration is that
we have to take into account all {\irs} arising from the unfolding
process and then use Eq.~\eqref{eq:folding_vecs} to reconstruct the
bandstructure from the ARPES data, if necessary.

\section{Applications}

\begin{figure}
 \begin{center}
  \includegraphics{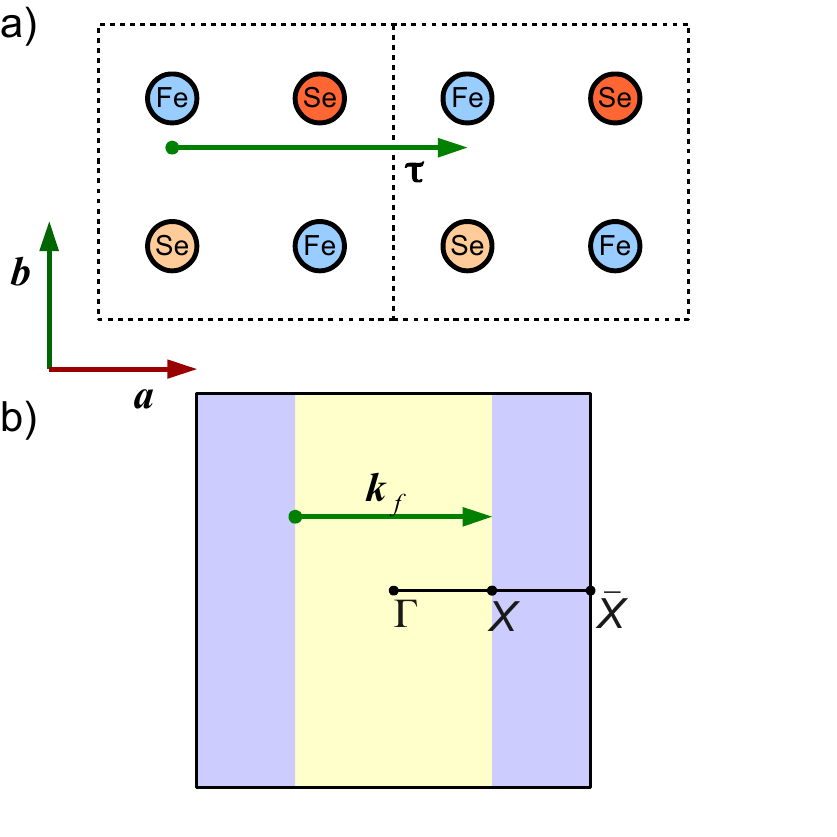}
 \end{center}
\caption{ (a) Projection on the $ab$-plane of a $2\times1$ supercell 
  of tetragonal FeSe.
The fractional translation $\fractrans$ along $\vect{a}$ is shown in 
green color. The iron atoms
lie in the same plane, parallel to the plane of the drawing, while the lighter
colored selenium atoms are vertically displaced above the plane, and the darker 
colored ones are below. (b) $k_z=0$ plane of the Brillouin zone corresponding to
the single unit cell of FeSe. The yellow filling marks the Brillouin zone
corresponding to the supercell. The folding vector $\kvec_f$ is shown in green
color.}\label{fig:fese_doubled_cellbz}                                 
\end{figure}

Iron-based superconductors provide an excellent playground for the unfolding
method presented here. The crystal structures of iron pnictides and iron
chalcogenides consist of layers of Fe atoms tetrahedrally coordinated by the
pnictogen/chalcogen atoms as shown in Fig.~\ref{fig:fese_doubled_cellbz}~(a). The
various compounds may show differences in the stacking sequence of the iron
pnictogen/chalcogen layers, as well as in the  composition of the spacer layers.
For most of the iron pnictide/chalcogenide families, a minimal translationally
invariant unit cell consists of two iron and two pnictogen/chalcogen atoms. This
unit cell can further be reduced by considering the glide-mirror operations,
which combine the translations between the nearest-neighbor iron atoms with
reflections in the $xy$-plane, thus mapping two translationally inequivalent
iron and pnictogen/chalcogen sites into each other.

In the following we shall consider four representative unfolding examples.
In the first case we will apply translational unfolding on a $2\times1$ 
supercell of FeSe where the translational symmetry is kept in the
supercell.  In the second case we will apply translational 
unfolding on a P-doped CaFe$_2$As$_2$ $2\times2$ 
supercell (Ca$_4$Fe$_8$As$_7$P)
where the translational symmetry
has been broken by the substitution of one As by P. In the third
case we unfold the 16-band tight-binding model ($2\times5$ Fe bands
and $2\times3$ Se bands) for FeSe at 10 GPa to an 8-band tight-binding model.
At this pressure, the structure shows important dispersion along 
$k_z$ and it allows for an analysis of the unfolding procedure in all
three directions.
Finally in the fourth case we apply unfolding of the two-iron unit cell
to the one-iron unit cell representation in the specific case of
the body centered space group $I 4/mmm$. With this last example
we want to show that the unfolding procedure is independent of whether
the space group is body centered or not.

\begin{figure*}
 \begin{center}
  \includegraphics[width=\textwidth]{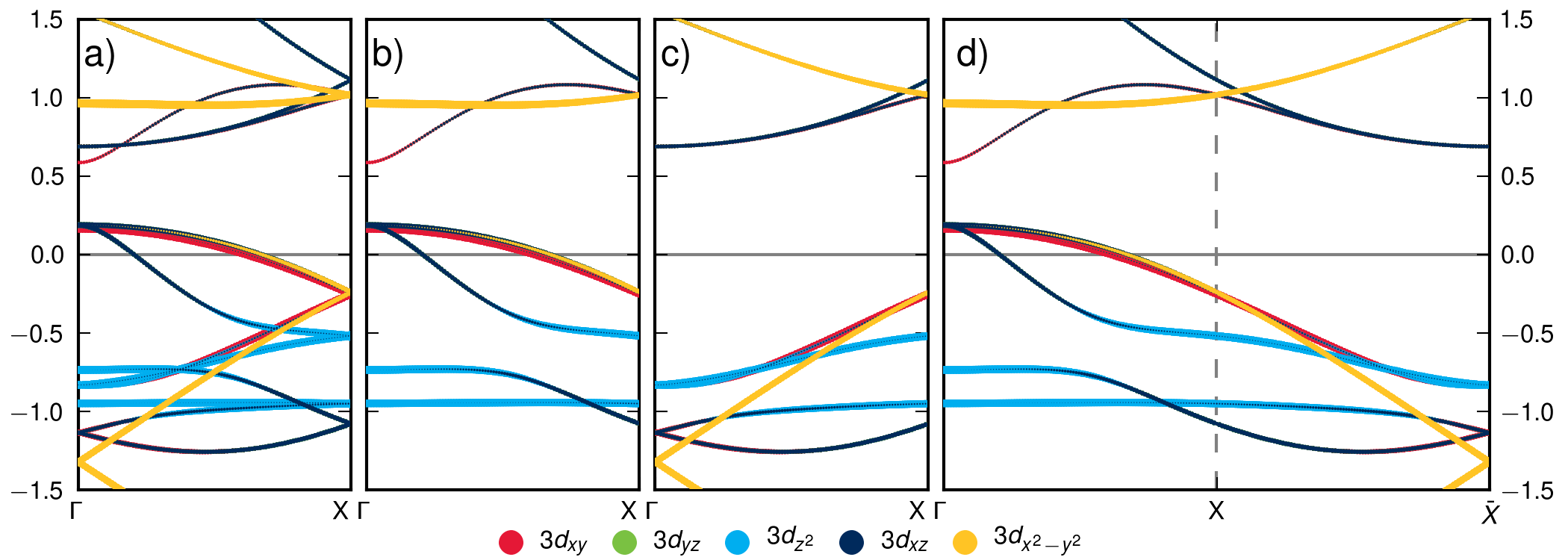}
 \end{center}
\caption{Unfolding of the tetragonal FeSe doubled along the $\vect{a}$-axis.
(a) The folded bandstructure. (b) Projection of bands onto $\irrep{\kvec,0}$.
(c) Projection of bands onto $\irrep{\kvec,0}$. (d) Unfolded picture. \Ir~
$\irrep{\kvec,0}$ is extended past the boundaries of the Brillouin zone by the
folding vector $\kvec_f=(\pi/a,0,0)$.}\label{fig:fese_doubled}
\end{figure*}

\subsection{ Translational unfolding of FeSe}

As a first example, we consider a simple case of translational
unfolding, where the translation group $\gtrans$ is expanded by fractional
translations $\cosopi=(1|\fractransi)$.  We define a
 $2\times 1$ supercell of  tetragonal FeSe (Fig.~\ref{fig:fese_doubled_cellbz} (a))
by  doubling the unit cell along the $\vect{a}$ axis of FeSe.
In Fig.~\ref{fig:fese_doubled} (a) we show the corresponding 
supercell bandstructure
 along the path $\Gamma-X$ (Fig.~\ref{fig:fese_doubled_cellbz} (b)).
 In order to unfold the bands,
 we employ the fractional translation $\fractrans=\vect{a}$ which
is an additional
symmetry that the supercell has on top of the
translational symmetry $\gtrans$. With this, the factor
group $\factgroup$ is isomorphic to the cyclic group of order two, with the
generator $(1|\fractrans)$ and from every {\ir} $\gtrans$, two {\irs}
with $\alpha=0$ and $\alpha=1$ are induced, so that the generator
Eq.~\eqref{eq:allowed_reps} is represented by
\begin{equation}\begin{split}\label{eq:trans_unfolding_gen}
 \irrepel{\kvec,0}{(1|\fractrans)} &= \exp(-\ii\kvec\cdot\fractrans) \\
 \irrepel{\kvec,1}{(1|\fractrans)} &= -\exp(-\ii\kvec\cdot\fractrans) \\
\end{split}\end{equation}
The resulting band projections onto the {\irs} are shown in Fig.
\ref{fig:fese_doubled} (b) and (c) while the unfolded picture, where the {\ir}
$\irrep{\kvec,0}$ is extended outside the supercell BZ is shown in Fig.
\ref{fig:fese_doubled} (d). Evidently $\irrep{\kvec+\kvec_f,0} = \irrep{\kvec,
1}$, with $\kvec_f = (\pi/a,0,0)$.

\subsection{Translational unfolding of P-doped CaFe$_2$As$_2$}

The unfolding shown on Fig.~\ref{fig:fese_doubled} is perfect, because the
fractional translation $(1|\fractrans)$ is an exact symmetry of the supercell
and every band will belong to only one of the {\irs} of $\sgroup$. In a more
realistic case, where the operations $\cosopi$ are only approximate symmetries,
the bands will have nonzero projections onto multiple {\irs}, although usually,
one of the {\irs} will be dominantly present in every band. Such a situation
occurs, for example, when studying doped compounds. Here we have chosen to
investigate the phosphorus doped \Ca.

It is well known that when pressure is applied on {\Ca}, it undergoes a
magneto-structural phase transition from a magnetically ordered orthorhombic
phase to a non-magnetic, collapsed tetragonal phase \cite{Torikachvilli2008,
Kreyssig2008}. In previous studies~\cite{Zhang2009,Tomic2012,Tomic2013}, we
simulated the application of pressure under different conditions by means of
density functional theory calculations and were able to predict the appearance
of the collapsed tetragonal phase at a critical pressure which is accompanied by
the disappearance of the Fermi surface pockets centered around the $\Gamma$
point. This feature has been recently confirmed by angle resolved photoemission
experiments~\cite{Dhaka2014,Gofryk2014}. %(RV: here include the Dhaka paper as well as
%Phys. Rev. Lett. 112, 18640).

An orthorhombic to collapsed tetragonal phase transition in {\Ca} can also be
induced by chemical pressure. For example, substitutional doping of phosphorus
into the arsenic sites causes \Ca~ to enter the collapsed tetragonal phase at a
doping level of around 5\%~\cite{Kasahara2011}. In order to fully understand how
chemical pressure is related to the application of physical pressure, we have
performed a sequence of full structural relaxations of P-doped \Ca. For the
different doping levels we have considered supercells of various sizes. Our
density functional theory 
calculations predict that P-doped {\Ca} undergoes an orthorhombic to collapsed
tetragonal phase transition for a doping between 9.375\% and 12.5\% in good
agreement with the experimental observations~\cite{Kasahara2011}. In
order to analyze the electronic structure in the collapsed tetragonal phase of
P-doped \Ca, we have to perform the unfolding of the bandstructure.

\begin{figure}
\begin{center}
\includegraphics[width=0.5\textwidth]{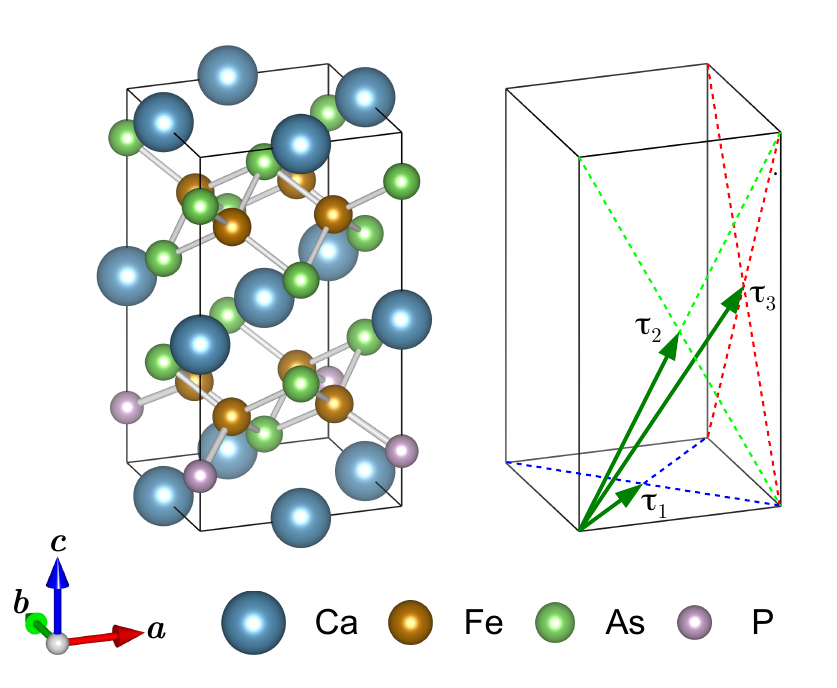}
\end{center}
\caption{The unit cell of \PCa. The fractional translations $\fractransi$ are
shown in green color.}\label{fig:p_doped_ca_cell}
\end{figure}

\begin{figure*}
 \begin{center}
  \includegraphics[width=\textwidth]{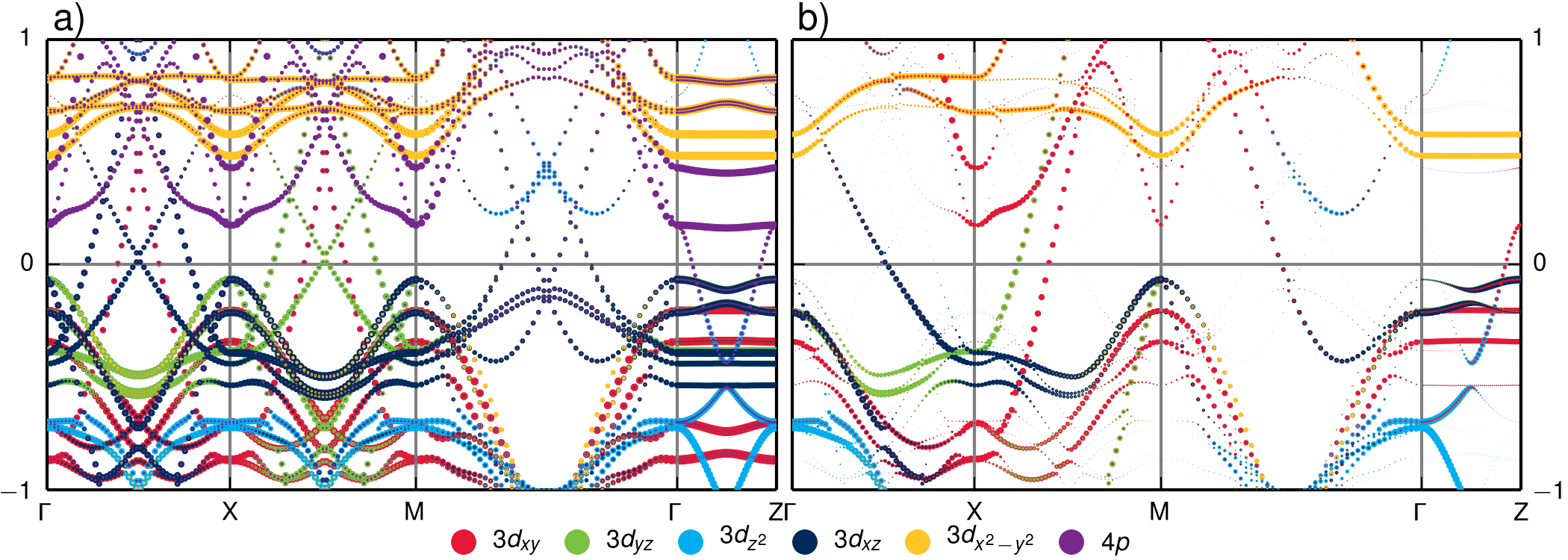}
 \end{center}
 \caption{Unfolding of {\PCa} bandstructure. (a) Bandstructure of the
   {\PCa} supercell. (b) The unfolded bandstructure obtained from
   $\irrep{\kvec,0}$.}
\label{fig:p_doped_ca_bands}
\end{figure*}

Relaxed {\PCa} has an orthorhombic unit cell where the phosphorus atom
is breaking the translational symmetry as shown in
Fig.~\ref{fig:p_doped_ca_cell}.  The unit cell of {\PCa} is a
supercell consisting of four primitive unit cells of {\Ca}. This unit
cell contains a total of eight iron atoms. The corresponding
fractional translations are $\fractrans_1 = (\vect{a} + \vect{b})/2$,
$\fractrans_2 = (\vect{a} + \vect{c})/2$ and $\fractrans_3 =
\fractrans_1 + \fractrans_2$, where $\vect{a}$, $\vect{b}$ and
$\vect{c}$ are the unit vectors of the supercell, as shown in
Fig.~\ref{fig:p_doped_ca_cell}.  These fractional translations map the
two translationally inequivalent iron atoms to the eight iron atoms of
the supercell.

The corresponding bandstructure is shown in
Fig.~\ref{fig:p_doped_ca_bands} (a).  The bandstructure is calculated
along the path given by $(0,0,0) - (2\pi/a,0,0) - (2\pi/a,2\pi/a,0) -
(0,0,0) - (0,0,2\pi/c)$. Four \irs~can be induced. We will select the
{\ir} $\irrep{\vect{k},0}$ and extend it to the unfolded Brillouin
zone. The resulting unfolded bandstructure, obtained by extending the
\ir~ $\irrep{\kvec,0}$ is shown in Fig.~\ref{fig:p_doped_ca_bands}
(b). Despite the fact that the fractional translations $\fractransi$
are not the exact symmetries of {\PCa}, the band projections onto
different {\irs} are still mostly orthogonal, having relatively clean
unfolded bands as a result. This allows us to clearly see the
disappearance of the hole pockets centered around $\Gamma$, since the
set of three hole $t_{2g}$ bands is pushed below the Fermi level by
around 0.2~eV. 
%In Fig.~\ref{fig:p_doped_ca_bands} (c) and (d)
%we also show the Fermi surface cuts at $k_z=0$.
Comparison of the unfolded bands to the bandstructure
of the collapsed tetragonal phase of {\Ca} under
pressure~\cite{Tomic2012,Dhaka2014} confirms that phosphorus doping
and application of hydrostatic pressure affect the structural and
electronic properties of {\Ca} in a remarkably similar way.

\subsection{Unfolding from a 16-band to an 8-band tight-binding model for
  FeSe under pressure}

We will now demonstrate the use of Eq.~\eqref{eq:unfolded_hoppings}
and Eq.~\eqref{eq:unfolded_hoppings_final} for the unfolding of
tight-binding models. We consider as a test system FeSe at 10~GPa with
significant dispersion in all three directions. The crystal structure
has been obtained from ab-initio simulations of hydrostatic pressure
at 10~GPa~\cite{Tomic2012,Tomic2013}. We have used the projective
Wannier functions as implemented in the FPLO code~\cite{Koepernik1999}
for the derivation of the 16-band tight-binding model which consists
of five $3d$ orbitals per iron site, and three $4p$ orbitals per
selenium site. The structure under 10~GPa of hydrostatic compression
is chosen because the three dimensional character of the Fermi surface
is more pronounced compared to ambient pressure.

As previously mentioned, the two translationally nonequivalent iron
sites can be mapped onto each other with the help of the glide mirror
operations $\cosopi = \seitzop{\fractransi}{\mat{\sigma}_z}$ with
$i=1,2$. The fractional translations $\fractransi$ connect the nearest
neighbor iron atoms as shown in Fig \ref{fig:fese_doubled_cellbz} (a),
while $\mat{\sigma}_z$ is a reflection in the $xy$-plane. To unfold we
can choose one of the $\cosopi$ operations and then induce the {\irs}
of $\sgroup=\gtrans\cup\gtrans\cosopi$. In accordance with
Ref.~\onlinecite{Andersen2011}, we call $\sgroup$ the glide-mirror
group. Since the factor group is of index two, two one-dimensional
{\irs} will be induced in the $k_z=0$ plane of the Brillouin
zone. Because the electronic dispersion in FeSe is weaker along the
$\kvec_z$ axis, we can expect that {\irs} induced in the $k_z=0$ plane
will give adequate unfolding across the rest of the Brillouin zone.

In these two {\irs}, the glide mirror operation will have the same
representation as did the fractional translation in
Eq.~\eqref{eq:trans_unfolding_gen}. However, what differentiates the
case of the glide-mirror unfolding from the purely translational
unfolding is the orbitally selective action of the matrices
$\mat{W}(\cosopi)$ in Eq.~\eqref{eq:projectors}. Namely, in the case of
translational unfolding, matrices $\mat{W}(\cosopi)$ act equally on
all orbitals, while in the case of glide-mirror unfolding, they act
differently, depending on whether the orbitals are symmetric or
antisymmetric with respect to the reflections in the $xy$-plane. For
example, the $3d_{z^2}$ orbital will stay invariant, while $3d_{xz}$
will pick up a minus sign under the action of $\mat{\sigma}_z$.

\begin{figure*}
 \begin{center}
  \includegraphics[width=\textwidth]{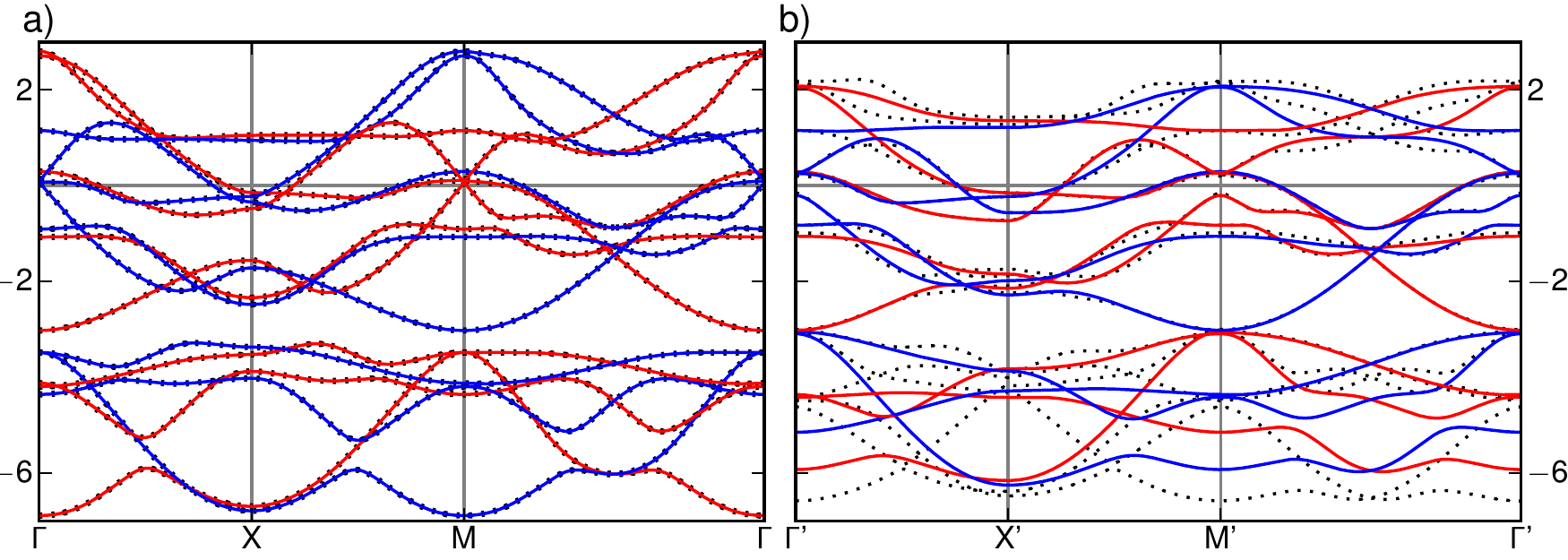}\vspace{-0.15cm}
  \includegraphics[width=\textwidth]{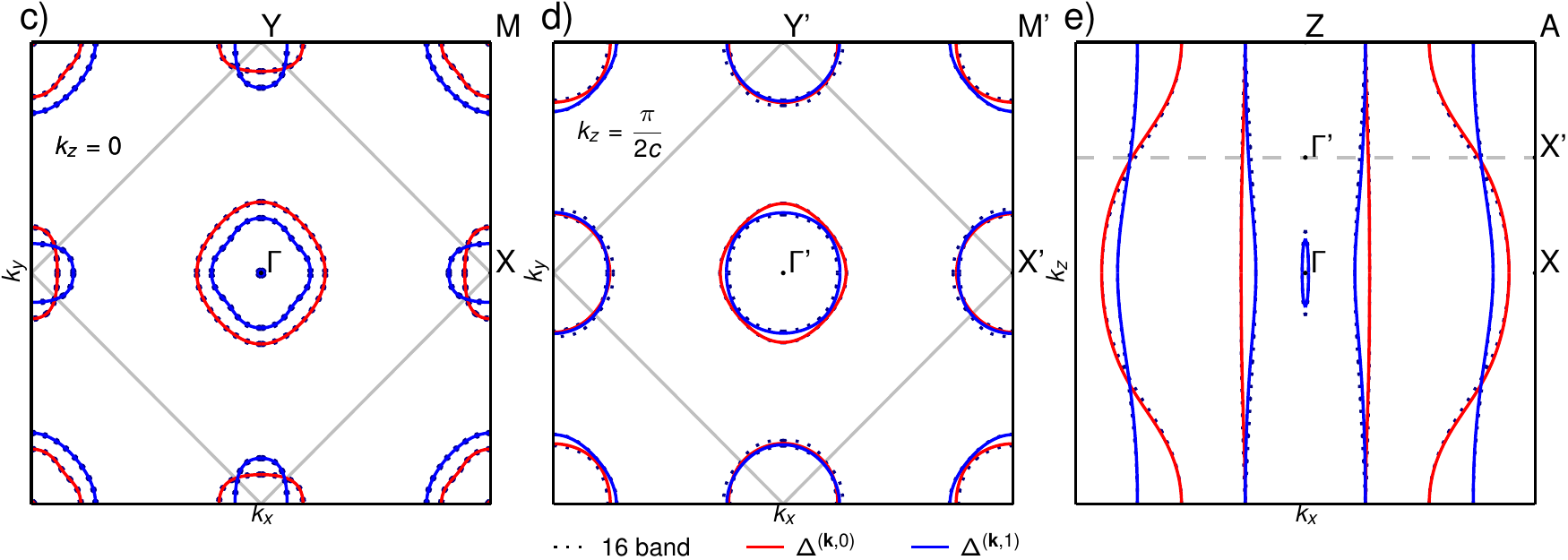}
 \end{center}
 \caption{Unfolding of the 16-band tight-binding model of FeSe with
   the help of glide-mirror operations. The 16-band result is shown
   with dots, the two 8-band results with lines. (a) Bandstructure
   along the path in the $k_z=0$ plane of the one iron equivalent
   Brillouin zone. (b) Bandstructure along the path parallel to the
   path shown in (a), but shifted along the $\kvec_z$-axis by
   $\pi/2c$. (c) Fermi surface cut at $k_z=0$ plane of the one-iron
   equivalent Brillouin zone of FeSe. The solid gray line shows the
   boundary of the two-iron Brillouin zone. (d) Fermi surface cut at
   $k_z=\pi/2c$ plane of the one-iron equivalent Brillouin zone of
   FeSe. (e) Fermi surface cut at $k_y=0$ plane of the one-iron
   equivalent Brillouin zone of FeSe. The dashed gray line shows the
   location of $k_z=\pi/2c$.} \label{fig:fese_tb_model}
\end{figure*}

We have used Eq.~\eqref{eq:unfolded_hoppings} to create the two sets
of hopping energies, corresponding to two induced {\irs}. These two
sets correspond to the same 8-band tight-binding model, up to the
unitary transformation. The bandstructure calculated from the 16-band
and two 8-band tight-binding models along the path in the $k_z=0$
plane of the Brillouin zone is shown in Fig.~\ref{fig:fese_tb_model}
(a). The corresponding Fermi surface slice in the $k_z=0$ plane is
shown in Fig.~\ref{fig:fese_tb_model} (c). In accordance with
Eq.~\eqref{eq:folding_vecs} and the electronic structure shown in
Fig.~\ref{fig:fese_tb_model} the folding vector is $\kvec_f^{01} =
(\pi,\pi,0)$ with respect to one iron Brillouin zone (as a convention,
we will always specify folding vectors with respect to the unfolded
Brillouin zone). It is evident that the unfolding to the 8-band model
is perfect in the $k_z=0$ plane. Since the $k_z=\pm\pi/c$ planes are
also the high-symmetry planes for the reflections in the $xy$-plane,
the unfolding will be perfect there too. We can thus expect the
largest deviations from the perfect unfolding around the
$k_z=\pm\pi/2c$ plane.  This can be seen in the bandstructure shown in
Fig.~\ref{fig:fese_tb_model} (b), taken along the path shown in
Fig.~\ref{fig:fese_tb_model} (a) shifted by $(0,0,\pi/2c)$. The
deviations of the unfolded bands are evident. However, the deviations
of the top ten bands, which are the bands dominated by the $3d$
orbital character, are much smaller than in the bands dominated by the
$4p$ orbital character. This is a consequence of the crystal
structure; the iron atoms are stationary under the action of
$\mat{\sigma}_z$, while the selenium atoms are not. Due to this
property, the Fermi surface can be unfolded almost exactly across the
entire Brillouin zone. The Fermi surface slice in the $k_z=\pi/2c$ is
shown in Fig.~\ref{fig:fese_tb_model} (d), while the vertical slice in
the $k_y=0$ plane is shown in Fig.~\ref{fig:fese_tb_model} (e). It is
remarkable that the full three dimensional structure of the innermost
Fermi surface pocket, centered at $\Gamma$, is retained with high
accuracy in the unfolded model, despite the fact that the underlying
unfolding symmetry is purely two-dimensional.

\subsection{Unfolding the bandstructure of {\Ca} to the one-iron
  equivalent Brillouin zone}

\begin{figure*}
 \begin{center}
  \includegraphics[width=\textwidth]{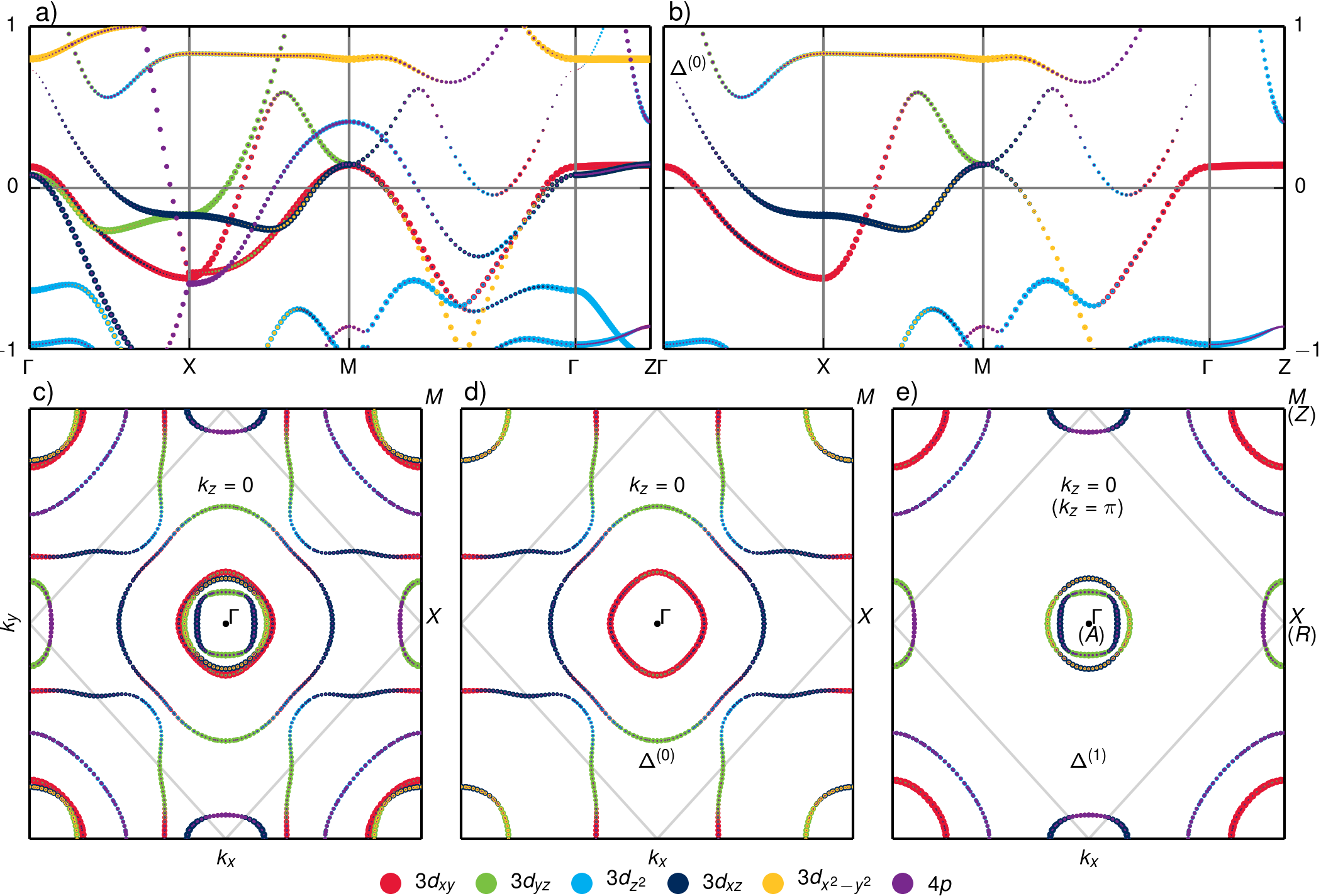}
 \end{center}
 \caption{Unfolding of {\Ca} bandstructure to one iron equivalent
   Brillouin zone.  (a) Bandstructure of {\Ca} in the two-iron
   Brillouin zone. (b) The unfolded bandstructure obtained from
   $\irrep{\kvec,0}$ in the one-iron equivalent Brillouin zone. (c) $k_z=0$
   cut of {\Ca} Fermi surface cut in the two-iron Brillouin zone. (d) 
   and (e) Projection of the Fermi surface obtained from $\irrep{\kvec,0}$ 
   and  $\irrep{\kvec,1}$, respectively. Labels enclosed in brackets on 
   panel (e) pertain to $k_z=\pi$ cut of the Fermi 
   surface.}\label{fig:cafe2as2_1fe_unfolding}
\end{figure*}

\begin{figure}
 \begin{center}
  \includegraphics[width=0.45\textwidth]{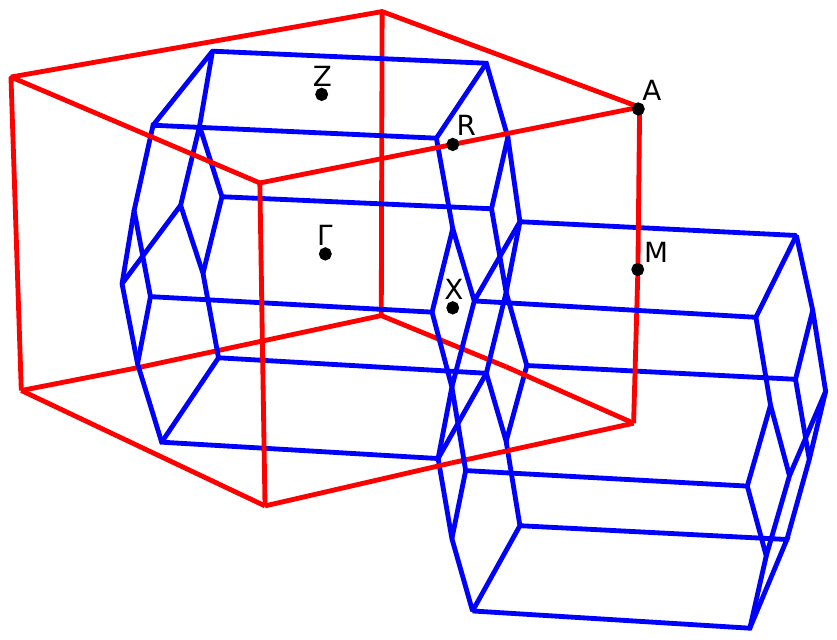}
 \end{center}
 \caption{One and two iron Brillouin zones for the $I4/mmm$
   spacegroup. Two adjacent Brillouin zones corresponding to the two
   iron primitive cell of $I4/mmm$ space group are shown in blue
   color. The simple tetragonal unit cell corresponding to the one
   iron unit cell is shown in red.}
  \label{fig:bz_drawing}
\end{figure}

Finally, we would like to point out that the glide-mirror group can
also be used to unfold the bandstructures of iron-based
superconductors with a centered unit cell, described by symmorphic
space groups.  In this particular example, we use the ambient pressure
structure of {\Ca} measured at a temperature $T=250$~K and described
by the body-centered spacegroup $I\,4/mmm$~\cite{Kreyssig2008}. {\Ca}
was selected because it features additional electron Fermi surface
pockets which make the one iron Fermi surface determination more
complicated.

The glide-mirror operations as well as the {\irs} are identical to the
case of FeSe. The main difference here is that the resulting folding
vector is $(\pi,\pi,\pi)$, corresponding to the unfolding onto the
simple tetragonal one iron unit cell. This result is not immediately
obvious, since we are only using the symmetry of the iron-arsenic
layer, and it is to be expected, in the na{\"i}ve picture, that the
resulting unfolding would result in the one iron body-centered unit
cell and the $(\pi,\pi,0)$ folding vector. However, due to the
body-centered symmetry the $(\pi,\pi,0)$ folding vector relates
$\Gamma$ and $Z$ points, as shown in Fig.~\ref{fig:bz_drawing}. In
addition, unlike $(\pi,\pi,\pi)$, the $(\pi,\pi,0)$ folding vector
does not satisfy Eq.~\eqref{eq:folding_vecs}, which underlines their
usefulness.

Fig.~\ref{fig:cafe2as2_1fe_unfolding}~(a) shows the bandstructure in
the two-iron equivalent Brillouin zone, while
Fig.~\ref{fig:cafe2as2_1fe_unfolding}~(b) shows the unfolded
bandstructure in the one-iron equivalent Brillouin zone obtained by
extending $\irrep{\kvec,0}$.  Evidently, the unfolding remains perfect
in the $k_z=0$ plane. However, this is no longer the case across the
entire Brillouin zone. For instance, along the $\Gamma-Z$ path weak
traces of bands from $\irrep{\kvec,1}$ can be observed. In comparison
to FeSe, in the vicinity of the Femi level, the {\Ca} bandstructure
features more orbital weight of arsenic $4p$ and some calcium $3d$
character which makes unfolding outside of the high-symmetry plane
less accurate. Nevertheless, we have enough information to clearly
discern the topology of the unfolded Fermi
surface. Figs.~\ref{fig:cafe2as2_1fe_unfolding}~(c)-(e) show the
$k_z=0$ cut of the {\Ca} Fermi surface. Because of the body-centered
arrangement of the Brillouin zones shown in Fig. \ref{fig:bz_drawing}
this is also the $k_z=\pi$ cut of the Fermi surface, offset by
$(\pi,\pi,0)$ and as a result the connectivity of the Fermi surface
elements can be deduced between the $k_z=0$ and $k_z=\pi$ planes.

\section{Discussion}

\subsection{One-iron vs. two-iron picture}

The above results have important implications for the one-iron vs. two-iron
discussion in the iron pnictide literature~\cite{Lin2011}. Since computation of
such properties like multiorbital pairing calculations~\cite{Graser2009} scale
as the sixth power of number of orbitals involved, it is important to keep the
models as simple as possible. In addition, careful consideration of symmetry has
very important consequences for the superconducting state \cite{Cvetkovic2014,Casula2013,Ku2014}.
For this reason, one needs to understand the exact conditions and the symmetry
context in which the one-iron model can be used.

When the unfolding is considered as a projection onto the irreducible subspace
of the glide-mirror group, some potentially important subtleties arise in
comparison to the conclusions drawn in Ref.~\onlinecite{Lin2011}. As long as the
electronic structure in the energy range of interest is dominated by the iron
orbitals and all dynamics under consideration involve at least approximate
glide-mirror symmetry, a one-iron tight-binding model can be used without
significant impact on the overall accuracy of the calculation. Furthermore, even
in the cases where the glide-mirror symmetry is not so favorable, a controlled
one-iron approximation can be made since the off-diagonal blocks in the
decomposition onto irreducible subspaces can provide an error estimate. In
addition, this shows that only the glide-mirror group can provide criteria that
unambiguously resolve Fermi surface elements in the process of unfolding to the
one-iron Brillouin zone.

Furthermore, our discussion of effects of high-symmetry elements of
the Brillouin zone on the interpretation of ARPES data shows that
under certain conditions it is possible to observe the spectral
function consistent with the one iron picture in $k_z=0$ or $k_z=\pi$
planes. This is corroborated by recent observations \cite{Kong2014} in
CsFe$_2$As$_2$ whose electronic structure is weakly dispersive in the
$k_z$ direction and the glide mirror unfolding can be accurately
extended across the entire Brillouin zone.

The unfolding process also offers a simple answer to the question why the
neutron scattering intensities seem to indicate a scenario consistent with the
one-iron picture~\cite{Lumsden2010,Park2010}. This can be naturally interpreted
as a consequence of the fact that neutron scattering intensities are
momentum-resolved in the high symmetry plane and  transitions between states
belonging to different irreducible subspaces are suppressed there.

\section{Summary}

In summary, we have demonstrated with a group-theoretical treatment of the
bandstructure that unfolding  can be understood as a projection onto induced
{\irs} of the supergroup of the original translation group. The unfolded
Brillouin zone arises as a consequence of the fact that different induced {\irs}
become identical when shifted by an appropriate vector in the Brillouin zone.
Due to the projective definition, the unfolding procedure can be generalized for
arbitrary quantities in reciprocal space. Also, the unfolding artifacts in the
cases where the unfolding is inexact arise because bands have nonzero
projections onto multiple {\irs}.

When point group operations are used, the unfolding is exact only in the
high-symmetry k-points of the Brillouin zone. It is nonetheless possible to
extend the unfolding to the entire Brillouin zone as long as the bandstructure
is dominantly dispersive only along the corresponding high-symmetry lines or
planes in the Brillouin zone. By making sure this constraint is satisfied, it is
possible to formulate tight-binding models of reduced dimensionality without 
loss of accuracy. In the cases where this is not completely possible, the
unfolding framework provides a systematic way to make controlled approximations
by projecting the relevant quantities into appropriate irreducible subspaces.

For FeSe under pressure, we have shown how an 8-band tight-binding model can be
constructed by unfolding the 16-band tight-binding model with the help of
glide-mirror operations. The resulting unfolded model produces the almost
exactly unfolded Fermi surface. This results from the fact that the Fermi
surface in FeSe is dominated by the iron orbital character. This is in fact the
most important requirement that needs to be fulfilled for the one iron model to be
an accurate representation of the physics of iron based superconductors. An additional
requirement stems from the fact that the one iron picture is formulated as an
irreducible representation, and as such for any realistic computation using the one
iron model, the off-diagonal elements of the involved observables, connecting the two
irreducible representations of the glide-mirror group need to be small compared
to the diagonal elements.

And finally, we show that careful interpretation of ARPES data in cases where
the direct comparison with density functional theory calculations is not
immediately obvious, requires consideration of possible effects of high
symmetry elements of the Brillouin zone as well as comparison with all {\irs}
arising from the unfolding process.

\begin{acknowledgments}
We would like to thank Doug J. Scalapino, Peter J. Hirschfeld and Lilia Boeri for
useful discussions and we gratefully acknowledge financial support by the
Deutsche Forschungsgemeinschaft through grant SPP 1458.
\end{acknowledgments}

\end{document}